\documentclass[aps,prd,amssymb,eqsecnum,showpacs]{revtex4}
\usepackage{graphicx}
\usepackage{psfrag}

\newcommand{\gz}{\mbox{\em \r{g}\hspace{0.1mm}}}  % Zero order approximation.
\newcommand{\Rz}{\mbox{\em \r{R}}}

\newcommand{\nablaz}{\nabla\hspace{-0.27cm}{}^{\mbox{\r{~}}}{}\hspace{-0.22cm}}

\newcommand{\Gammaz}{\Gamma\hspace{-0.20cm}{}^{\mbox{\r{~}}}{}\hspace{-0.18cm}}

\begin{document}

\title{Disembodied boundary data for Einstein's equations}

\author{Jeffrey Winicour${}^{1,2}$
       }
\affiliation{
${}^{1}$ Department of Physics and Astronomy \\
         University of Pittsburgh, Pittsburgh, PA 15260, USA \\
${}^{2}$ Max-Planck-Institut f\" ur
         Gravitationsphysik, Albert-Einstein-Institut, \\
	 14476 Golm, Germany
	 }

\begin{abstract}

A strongly well-posed initial boundary value problem based upon
constraint-preserving boundary conditions of the Sommerfeld type has been
established for the harmonic formulation of the vacuum Einstein's
equations. These Sommerfeld conditions have been previously presented in
a 4-dimensional geometric form. Here we recast the associated boundary
data as 3-dimensional tensor fields intrinsic to the boundary. This
provides a geometric presentation of the boundary data analogous to the
3-dimensional presentation of Cauchy data in terms of 3-metric and
extrinsic curvature. In particular, diffeomorphisms of the boundary data
lead to vacuum spacetimes with isometric geometries. The proof of
well-posedness is valid for the harmonic formulation and its
generalizations. The Sommerfeld conditions can be directly
applied to existing harmonic codes which have been used in simulating
binary black holes, thus ensuring boundary stability of the underlying
analytic system. The geometric form of the boundary conditions also
allows them to be formally applied to any metric formulation of
Einstein's equations, although well-posedness of the boundary problem is
no longer ensured.  We discuss to what extent such a formal application
might be implemented in a constraint preserving manner to 3+1
formulations,  such as the Baumgarte-Shapiro-Shibata-Nakamura system
which has been highly successful in binary black hole simulation.

\end{abstract}


\maketitle

\section{Introduction}

Previous work has established the strong well-posedness of the
initial-boundary value problem (IBVP) for Einstein's equations expressed
as a hyperbolic system in harmonic coordinates. The result was first
obtained using pseudo-differential theory, i.e. using a Fourier-Laplace
expansion, which established well-posedness in the generalized
sense~\cite{wpgs}. The result was subsequently obtained using standard
energy estimates~\cite{wpe,isol}. This places the IBVP on the same
analytic footing as the Cauchy problem, whose well-posedness was also
established using harmonic coordinates in the classic work of
Choquet-Bruhat~\cite{Choquet}. The geometric formulation of the boundary
conditions and boundary data for the IBVP is more complicated than for
the Cauchy problem. Recently, this boundary data was presented in a
4-dimensional geometric form~\cite{juerg}. Here we recast the boundary
data as 3-dimensional tensor fields intrinsic to the boundary, analogous
to the presentation of Cauchy data in terms of the 3-metric and extrinsic
curvature of the initial Cauchy hypersurface. The spacetime metric which
solves the harmonic IBVP with this data is uniquely determined up to a
diffeomorphism.

\begin{figure}[htb]
\begin{center}
\includegraphics[scale=.4]{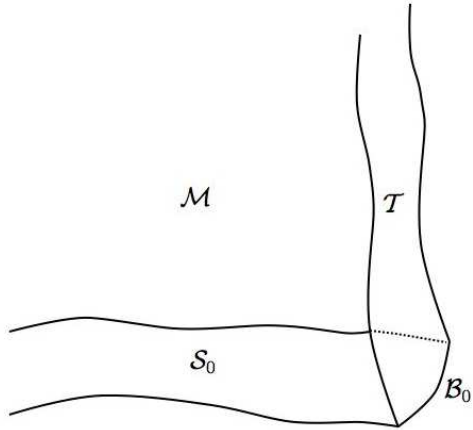}  
\caption{Data on the 3-manifolds ${\cal S}_0$ and ${\cal T}$, which
 intersect in the 2-surface ${\cal B}_0$, locally determine a solution
 in the spacetime manifold ${\cal M}$}.  
\end{center}
\label{fig:bound}
\end{figure}

In the Cauchy problem, initial data on a spacelike hypersurface ${\cal
S}_0$ determine a solution in the future domain of dependence ${\cal
D}^{+}({\cal S}_0)$ (which consists of those points whose past directed
characteristics all intersect ${\cal S}_0$).  In the IBVP, data on a
timelike boundary ${\cal T}$ which meets ${\cal S}_0$ in a surface 
${\cal B}_0$ are used to further extend the solution to the domain of
dependence ${\cal D}^{+}({\cal S}_0 \cup {\cal T})$. In practical
applications ${\cal B}_0$ is topologically a sphere surrounding some
system of interest but here we only consider the local problem in a
neighborhood of a point of intersection between the Cauchy hypersurface
and the boundary. For hyperbolic systems, the global solution in the
spacetime manifold ${\cal M}$ can be obtained by patching together local
solutions. The setting for this local problem is depicted in
Fig.~\ref{fig:bound}.

There are no natural boundaries for the gravitational field analogous to
the conducting boundaries that play a major role in electromagnetism.
Consequently, the IBVP for Einstein's equations only received widespread
attention after its importance to the artificial outer boundaries used in
numerical relativity was pointed out~\cite{stewart}. The first well-posed
IBVP was achieved for a tetrad formulation of Einstein's theory in terms
of a first differential order system which included the tetrad, the
connection and the curvature tensor as evolution fields~\cite{fn}. A
strongly well-posed IBVP was later established for the harmonic
formulation of Einstein's equations as a system of second order
quasilinear wave equations for the metric~\cite{wpgs,wpe}. Strong
well-posedness guarantees the existence of a unique solution which
depends continuously on both the Cauchy data and the boundary data. The
results were further generalized in~\cite{isol} to apply to general
quasilinear symmetric hyperbolic systems whose boundary conditions have a
certain hierarchical form. 

The initial data for the Cauchy problem can be formulated in terms of two
symmetric tensor fields $\tilde h_{ab}$ (a Euclidean 3-metric) and
$\tilde k_{ab}$ on a 3-manifold $\tilde{\cal S}_0$ subject to the
(Hamiltonian and momentum) constraints
\begin{equation}
0={}^{(3)} \tilde R +(\tilde k^a_a)^2
      -\tilde k_{ab}\tilde k^{ab} \quad      \label{eq:ham}
\end{equation}
and
\begin{equation}
 0 =\tilde \nabla_a (\tilde k^a_b -  \delta^a_b  \tilde k^c_c) ,
    \label{eq:mom}
\end{equation}    
where $\tilde \nabla_a$ is the covariant derivative and ${}^{(3)} \tilde
R$ is the curvature scalar associated with $ \tilde h_{ab}$. As
characterized in ~\cite{hawkel}, via an embedding in a 4-dimensional
manifold ${\cal M}$, the data   $\tilde h_{ab}$ and $\tilde k_{ab}$ on
the {\it disembodied} 3-manifold  $\tilde {\cal S}_0$ determine a
Lorentzian metric satisfying Einstein's equations which is unique up to
diffeomorphism  and whose restriction to the embedding ${\cal S}_0$ of
$\tilde{\cal S}_0$ gives rise to its intrinsic 3-metric $h_{ab}$ and
extrinsic curvature $ k_{ab}$.

Here we present an analogous result for the IBVP.  The difficulties
underlying the IBVP, which have most recently been discussed
in~\cite{juerg} and~\cite{hjuerg}, make the formulation of disembodied
boundary data more difficult than for the Cauchy problem. There are three
main complications.

\begin{itemize}

\item The first complication stems from a well-known property of the
IBVP for the flat-space scalar wave equation
$$
 (\partial_t^2 -\nabla^2)\Phi=0 
$$ in the region $x \le 0$.  Although the initial Cauchy data consist of
$\Phi|_{t=0}$ and $\partial_t \Phi|_{t=0}$, only half as much boundary
data can be freely prescribed at $x=0$, e.g the Dirichlet data
$\partial_t \Phi|_{x=0}$, or the Neumann data $\partial_x \Phi|_{x=0}$ or
the Sommerfeld data $(\partial_t + \partial_x) \Phi|_{x=0} $ (based upon
the derivative in the outgoing characteristic direction). For a given
physical problem, this implies that the boundary data cannot be
prescribed before the boundary condition is specified, i.e. the correct
boundary data depend upon the boundary condition, unlike the situation
for the Cauchy problem.  The analogue in the gravitational case is the
inability to prescribe both the metric and its normal derivative on a
timelike boundary, which implies that you cannot freely prescribe both
the intrinsic metric of the boundary and its extrinsic curvature. This
leads to a further complication regarding constraint enforcement on the
boundary, i.e. the Hamiltonian and momentum constraints (\ref{eq:ham})
and (\ref{eq:mom}) cannot be enforced directly because they couple the
metric and its normal derivative. Here we restrict our attention to
Sommerfeld boundary data which we prescribe in a constraint
free manner.

\item A Sommerfeld boundary condition for a metric component supplies the
value of the derivative $K^c \partial_c g_{ab}$ in an outgoing null
direction $K^c$. Such boundary conditions are very beneficial for
numerical work since they allow discretization error to propagate across
the boundary (whereas Dirichlet and Neumann boundary conditions reflect
the error and trap it in the numerical grid). However, the boundary does
not pick out a unique outgoing null direction at a given point (but,
instead, essentially a half null cone). This complicates the geometric
formulation of a Sommerfeld boundary condition. In addition, constraint
preservation does not allow specification of Sommerfeld data for all
components of the metric, as will be seen in formulating the Sommerfeld
conditions (\ref{eq:hk}) - (\ref{eq:hl}).

\item The third major complication arises from gauge freedom. In the
evolution  of the Cauchy data it is necessary to introduce a foliation of
the spacetime by Cauchy hypersurfaces ${\cal S}_t$, with unit timelike
normal $n_a$. The evolution of the spacetime metric
\begin{equation}
     g_{ab}=-n_a n_b +h_{ab}
     \label{eq:gdecom}
\end{equation}     
is carried out along the flow of an evolution vector field $t^a$ related
to the normal by the lapse $\alpha$ and shift $\beta^a$ according to
\begin{equation}
    t^a= \alpha n^a + \beta^a \, , \quad  \beta^a n_a =0.
    \label{eq:lapshif}
\end{equation}         The choice of foliation is part of the gauge
freedom in the resulting solution but does not enter into the
specification of the initial data. In the IBVP, the foliation is
unavoidably coupled with the formulation of the boundary condition. Thus
some gauge information must be incorporated in the formulation of the
boundary condition and boundary data.

\end{itemize}

In order to resolve these complications,  we include the specification of
a foliation ${\cal B}_t$ of the boundary ${\cal T}$  as part of the
boundary data. This supplies the gauge information which determines a
unique outgoing null direction for a Sommerfeld condition. In
Section~\ref{sec:main}, we formulate the 3-dimensional prescription of
the boundary data and the statement of our main result, a Theorem
establishing the existence of a solution satisfying a version of the
geometric uniqueness property proposed in~\cite{hjuerg}. The
3-dimensional description of the data in terms of scalar, vector and
tensor fields intrinsic to the boundary is quite abstract at this stage.
Their physical significance only becomes clear in
Section~\ref{sec:exist}, where the proof of the Theorem is given. The
proof  is based upon the well-posedness of the 4-dimensional geometrical
formulation of the harmonic IBVP given in~\cite{juerg}. Constraint
preservation is established by incorporating the harmonic conditions into
the boundary conditions.

The motivation for this work stems from the need for an improved
understanding and implementation of boundary conditions in the
computational codes being used to simulate binary black holes. We discuss
the applicability of our results to numerical relativity in
Sec.~\ref{sec:applic}.

Much of the presentation in the paper is coordinate independent and we use
Latin letters $(a,b,c,...)$ as abstract indices~\cite{wald} to denote the
types of vector and tensor fields and to indicate their manipulations. This
notation serves to describe either 4-dimensional tensor fields on the
spacetime ${\cal M}$ or 3-dimensional tensor fields intrinsic to ${\cal
S}_0$ or ${\cal T}$. When spacetime coordinates $x^\mu=(t,x^i)$ are
introduced, we use Greek letters $(\mu,\nu,\rho,...)$ to describe the
corresponding 4-dimensional tensor components and Latin letters
$(i,j,k,...)$ to denote the spatial components.

\section{Disembodied data for the IBVP}
\label{sec:main} 

We state our main result concerning Sommerfeld boundary data
for Einstein's equations.

\bigskip

{\bf Geometric Uniqueness Theorem:} Consider the 3-manifolds $\tilde
{\cal T}$ and $\tilde {\cal S}_0$ meeting in an edge $\tilde {\cal B}_0$.
On  $\tilde {\cal S}_0$ prescribe the smooth, symmetric tensor fields
$\tilde h_{ab}$  and $\tilde k_{ab}$, subject to the Hamiltonian and
momentum constraints and the condition that $\tilde h_{ab}$ be a
Riemannian metric . On $\tilde{\cal B}_0$ prescribe the smooth scalar
field $\tilde \Theta$. On $\tilde{\cal T}$ prescribe a smooth foliation
$\tilde{\cal B}_t$ parametrized by a scalar function $\tilde t$, where
(for convenience) $\tilde t=0$ on $\tilde {\cal B}_0$. In addition, on
$\tilde{\cal T}$, prescribe the scalar field $\tilde q$, the vector field
$\tilde q^a$ and the rank-2 symmetric tensor field $\tilde \sigma^{ab}$,
which are all smooth and vanish on $\tilde{\cal B}_0$.  Here the rank-2 
property of $\tilde \sigma^{ab}$ is defined with respect to the foliation
$\tilde{\cal B}_t$ by the requirement
\begin{equation}
       \tilde  \sigma^{ab} \partial_a \tilde t =0.
\end{equation}
Then, after embedding $\tilde{\cal S}_0 \cup \tilde {\cal T}$ as the
boundary ${\cal S}_0 \cup {\cal T}$ of a 4-manifold ${\cal M}$, as
depicted in Fig.~\ref{fig:bound}, this data provides Sommerfeld
boundary data for a vacuum spacetime, in a region including a neighborhood
of the embedded edge ${\cal B}_0$, which is unique up to diffeomorphism.

\bigskip

The geometrical interpretation of the data involves the metric of the
embedded spacetime, whose existence is the content of the Theorem. Before
proceeding to the proof in the next section, it is useful to supply some
intuitive meaning to the data. As in the Cauchy problem,  $\tilde
h_{ab}$  and $\tilde k_{ab}$ are identified with the 3-metric and the
extrinsic curvature of the embedding ${\cal S}_0$ of $\tilde {\cal S}_0$.
The scalar field $\tilde \Theta$ determines the hyperbolic angle
describing the initial velocity of the embedded boundary ${\cal T}$
relative to the inertial frame picked out by ${\cal S}_0$. The fields
$\tilde q$ and $\tilde q^a$ supply information  concerning the subsequent
dynamics of the boundary and its foliation. Together $\tilde q$ and
$\tilde q^a$ determine the components of a 4-vector $q^a$ describing the
curvature of the outgoing null geodesics normal in ${\cal M}$ to the
embedding ${\cal B}_t$ of $\tilde {\cal B}_t$. The field  $\tilde
\sigma^{ab}$ determines the optical shear $\sigma$ of the outgoing null
hypersurface through ${\cal B}_t$.

Note that the data contains no metric information about the embedded
boundary ${\cal T}$, not even that it is a timelike 3-manifold. Such
structure only emerges from the construction of a solution for the
spacetime metric. Thus fields derived algebraically from the metric, such
as the unit normal to the boundary, are to be considered as subsidiary
unknowns.

The requirement that the boundary data vanish at $\tilde {\cal B}_0$
stems from its interpretation relative to the initial Cauchy data.  This
requirement ensures the continuity of the resulting spacetime metric and
its first derivatives. The full compatibility conditions necessary for a
$C^\infty$ spacetime metric involve enforcing the Einstein equations and
their derivatives on ${\cal B}_0$. It is not clear how to implement $C^\infty$
compatibility conditions in terms of disembodied data. 

\section{The existence of a geometrically unique solution}
\label{sec:exist}

Our task is to show that harmonic evolution of the data prescribed in the
Geometric Uniqueness Theorem of Sec.~\ref{sec:main} locally determine a
vacuum spacetime which is unique up to diffeomorphism. We embed
$\tilde{\cal S}_0$, $\tilde {\cal T}$ and $\tilde {\cal B}_0$ in a
4-manifold ${\cal M}$ with boundary consisting of the corresponding
pieces ${\cal S}_0$, ${\cal T}$ and ${\cal B}_0$. The unknown is a
spacetime metric $g_{ab}$ on ${\cal M}$ which satisfies Einstein's
equations and is determined up to a diffeomorphism by the data specified
in the Theorem.  We follow the construction given in~\cite{juerg}.

The first step is to give initial data for  $g_{ab}$.   Let $h_{ab}$ and
$k_{ab}$ be the fields induced by $\tilde h_{ab}$ and $\tilde k_{ab}$ on
${\cal S}_0$; $\Theta$ be the field induced by $\tilde \Theta$ on ${\cal
B}_0$; and $\hat q$,  $\hat q^a$ and $\hat \sigma^{ab}$ be the fields
induced by $\tilde q$, $ \tilde q^a$ and $\tilde \sigma^{ab}$ on ${\cal
T}$. Let ${\cal B}_t$ be the foliation of ${\cal T}$ corresponding to the
embedding of $\tilde {\cal B}_t$, where $t$ is the parametrization
induced by $\tilde t$. On ${\cal S}_0$ prescribe a transverse field $n^a$
and construct the Lorentzian metric $ g^{ab}=-n^a n^b +h^{ab} $, so that
$n^a$ is the future directed unit normal to ${\cal S}_0$. Require that
$n^a$ satisfy $n^a N_a |_{{\cal B}_0} =\sinh \Theta$, where $N_a$ is the
outward unit normal to ${\cal T}$. 

We introduce the boundary decomposition of the metric 
\begin{equation}
      g_{ab} =N_a N_b -T_a T_b +Q_{ab},
      \label{eq:bdecom}
\end{equation}
where $T_a$ is the future directed unit normal in ${\cal T}$ to ${\cal
B}_t$. (Here the boundary fields $N_a$, $T_a$ and $Q_{ab}$ are unknowns
subsidiary to $g_{ab}$.) This leads to an orthonormal tetrad
$(T^a,N^a,Q^a,\bar Q^a)$ on ${\cal T}$, where $Q^a$ is a complex null
vector tangent to ${\cal B}_t$ with normalization
\begin{equation}
  Q_{ab}= Q_{(a}\bar Q _{b)}\, , 
  \quad Q^a \bar Q_a=2 \, , \quad  Q^a Q_a=0.
\label{eq:qnorm}
\end{equation}
(The tetrad is unique up to the spin freedom $Q^a \rightarrow e^{i\theta}
Q^a$ which does not enter our construction in any essential way.) 
Uniquely associated with this tetrad (independent of the choice of $Q^a$)
are the outgoing and ingoing null vector fields $K^a=T^a+N^a$ and
$L^a=T^a-N^a$, respectively, which lie in the null directions normal to
${\cal B}_t$. They form a null tetrad $(K^a,L^a,Q^a,\bar Q^a)$ with
metric decomposition
\begin{equation}
  g_{ab} = - K_{(a}L _{b)}+Q_{(a}\bar Q _{b)}.
\end{equation}

Next we tie down the gauge freedom by introducing an evolution field
$t^a$ on ${\cal M}$. We require that $t^a$ be tangent to ${\cal T}$ and
that it generate the foliation ${\cal B}_t$ according to 
\begin{equation}
               {\cal L}_t t =1,
\label{eq:fol}               
\end{equation}
where $ {\cal L}_t$ is the Lie derivative with respect to $t^a$. We
extend $t^a$ to ${\cal M}$ such that it generates a foliation ${\cal
S}_t$, with parametrization satisfying (\ref{eq:fol}) and unit normal
$n^a$. The gauge freedom is then pinned down by introducing spatial
coordinates $x^i$ on ${\cal S}_0$ and extending them to ${\cal M}$
according to
\begin{equation}
               {\cal L}_t x^i=0.
\label{eq:xi}               
\end{equation}
The scalars $x^\mu = (t,x^i)$ serve as coordinates for ${\cal M}$ which
are adapted to the evolution. Note that $t^a$ and the adapted coordinates
$x^\mu$ are explicitly constructed fields on ${\cal M}$ with no metric
properties. In the relationship (\ref{eq:lapshif}) between $t^a$ and the
unit normal $n^a$ to ${\cal S}_t$, it is $n^a$ which contains metric
information and is a subsidiary unknown.

As part of the initial data, we prescribe ${\cal L}_t n^a$ on
${\cal S}_0$. Along with the initial choice of $n^a$, $h_{ab}$ and
$k_{ab}$,  this determines the initial data $g_{ab}|_{t=0}$ and
${\cal L}_t g_{ab}|_{t=0}$. We use the evolution field $t^a$ to provide
a background metric ${\gz}_{ab}$ on ${\cal M}$ which is uniquely and
geometrically determined by the Lie transport of the initial data
according to
\begin{equation}
 {\gz}_{ab}|_{t=0}=g_{ab}|_{t=0} \, , \quad
   {\cal L}_t  {\gz}_{ab}|_{t=0} ={\cal L}_t  g_{ab}|_{t=0}\, ,   \quad
   {\cal L}_t  {\cal L}_t {\gz}_{ab} =0.
     \label{eq:gt0}
\end{equation}
In the coordinates $x^\mu=(t,x^i)$ adapted to the evolution, this
reduces to
\begin{equation}
 {\gz}_{\mu\nu}=g_{\mu\nu}|_{t=0} +t (\partial_t g_{\mu\nu}|_{t=0}).
     \label{eq:g0}
\end{equation}

The connection $\nablaz_a$ and curvature tensor $\Rz{}^d{}_{cab}$
associated with the background $\gz_{ab}$ have the same transformation
properties as the corresponding quantities $\nabla_a$ and $R^d{}_{cab}$
associated with $g_{ab}$. In particular, the difference $\nabla_a
-\nablaz_a$ defines a tensor field $C^d_{ab}$ according to
\begin{equation}
  (\nabla_a -\nablaz_a) v^d = C^d_{ab} v^b ,
  \label{eq:cconn}
\end{equation}
for any vector field $v^b$. In terms of the (nonlinear) perturbation
\begin{equation}
    f_{ab}=g_{ab}-\gz_{ab}
\label{eq:f}
\end{equation}
of the metric from the background, we have
\begin{equation}
  C^d_{ab}
   = \frac{1}{2} g^{dc}\left( \nablaz_{a} f_{bc} 
     + \nablaz_{a} f_{bc} - \nablaz_c f_{ab} \right).
     \label{eq:chris}
\end{equation}     

We take $ f_{ab}$ to be the evolution variable for solving Einstein's
equations. Since $\gz_{ab}$ is explicitly known, a solution for $ f_{ab}$
is equivalent to a solution for $g_{ab}$. By construction, the initial data
for $f_{ab}$ is homogeneous, i.e.
\begin{equation}
    f_{ab}|_{t=0}={\cal L}_t  f_{ab}|_{t=0} =0.
\label{eq:fdata}
\end{equation}
The boundary data on ${\cal T}$ consist of the vector field
\begin{equation}
    q^a = \hat q N^a + \hat q^a 
    \label{eq:qa}
\end{equation}
and the tensor field
\begin{equation}
      \sigma^{ab} = \hat  \sigma^{ab} - \frac{1}{2} Q^{ab} Q_{cd}\hat  \sigma^{cd}.
\end{equation}
Here, and elsewhere, the physical metric $g_{ab}$ is used to raise and
lower indices and to normalize the tetrad vectors.

Now, just as the Cauchy data $h_{ab}$ and $k_{ab}$ must be identified as the
intrinsic metric and extrinsic curvature of ${\cal S}_0$  in order to construct
a solution of the initial value problem, we give a geometric identification of
the boundary data. We identify this data as the geodesic curvature and shear,
relative to their background metric values, of the outgoing null vector $K^a$ on
${\cal T}$ according to the formulae
\begin{eqnarray}
     q^a &=& K^b (\nabla_b - \nablaz_b) K^a,
     \label{eq:qdata} \\
       \sigma^{ab} &=& \frac{1}{2}(Q^{ac}Q^{bd}- \frac{1}{2} Q^{ab}Q^{cd})(\nabla_c - \nablaz_c) K_d.
       \label{eq:shtendata}
\end{eqnarray}
The last equation can be re-expressed in the spin-weight-2 form
\begin{equation}
 \sigma:= Q^a Q^b \sigma_{ab} = 
     \frac{1}{2}Q^a Q^b(\nabla_a - \nablaz_a) K_b.
        \label{eq:shdata}
\end{equation}
The use of the shear in posing geometrical boundary conditions for the
harmonic formulation was suggested earlier in~\cite{ruiz}.

Using (\ref{eq:cconn}) and (\ref{eq:chris}), we recast (\ref{eq:qdata})
and (\ref{eq:shdata}) as Sommerfeld boundary conditions which determine
the components of the outgoing null derivatives  $K^a \nablaz_a f_{bc}$
according to
\begin{eqnarray} 
     \frac{1}{2} K^b K^c K^a 
      \nablaz_a f_{bc}  &=& q^a K_a ,
      \label{eq:ak}   \\
    (Q^b K^c K^a
       -\frac {1}{2} K^b K^c Q^a) \nablaz_a f_{bc} &=& q^a Q_a    ,
   \label{eq:aq}\\ 
     (L^b K^c K^a  
   -\frac {1}{2} K^b K^c L^a)
    \nablaz_a f_{bc}&=& q^a L_a ,
  \label{eq:al} \\
          (\frac {1}{2}Q^b Q^c K^a
       - Q^b K^c Q^a) \nablaz_a f_{bc} &=& 2\sigma .
   \label{eq:sigqq}       
\end{eqnarray}

In addition to these six Sommerfeld conditions, we impose the four
additional boundary conditions that  ${\cal C}^d: =g^{ab}  C^d_{ab}=0$
on  ${\cal T}$, i.e. the harmonic constraints. In terms of the null
tetrad decomposition, they take the Sommerfeld form
\begin{eqnarray} 
 -2{\cal C}^a K_a =\left( Q^b\bar{Q}^c  K^a+  K^b K^c L^a
            - K^b \bar{Q}^c Q^a  -  K^b Q^c \bar{Q}^a \right) 
	    \nablaz_a f_{bc}  &=& 0 , 
\label{eq:hk}  \\
 -2{\cal C}^a Q_a  =\left(  L^b Q^c K^a + K^b Q^c  L^a
            - K^b L^c  Q^a+Q^b Q^c  \bar{Q}^a \right)
	    \nablaz_a f_{b c} &=& 0,
\label{eq:hq} \\
 - 2{\cal C}^a L_a  = \left( L^b L^c  K^a+  Q^b \bar{Q}^c L^a
            - \bar{Q}^b L^c Q^a-  Q^b L^c \bar{Q}^a \right)
	    \nablaz_a f_{bc}&=& 0.
 \label{eq:hl}
\end{eqnarray}

Together, (\ref{eq:ak}) - (\ref{eq:hl}) provide Sommerfeld boundary
conditions for the  components of $K^a \nablaz_a f_{bc}$ in the
sequential order $(KK),(QK),(LK),(QQ),(Q\bar Q),(LQ),(LL)$ in terms of
the boundary data and the derivatives of preceding components in the
sequence.  Such a hierarchy of  Sommerfeld boundary conditions satisfy
the requirements of Theorem 1 of~\cite{isol} which establishes a
strongly  well-posed IBVP for a quasilinear hyperbolic system.

In order to apply this theorem, we reduce Einstein's equations to a
quasilinear wave system by modifying the Einstein tensor by the harmonic
constraints
\begin{equation}
  { \cal C}^d= g^{ab} C^d_{ab}
   = \frac{1}{2} g^{ab}g^{dc}\left( \nablaz_{a} f_{bc} 
     + \nablaz_{a} f_{bc} - \nablaz_c f_{ab} \right)
     \label{eq:harm}
\end{equation}
according to
 \begin{equation}
     E^{ab}:= G^{ab} -\nabla^{(a}{\cal C}^{b)} 
       +\frac{1}{2}g^{ab}\nabla_d{\cal C}^d .
       \label{eq:reduced}
\end{equation}
We eliminate the coordinate freedom, up to our choice of evolution
field $t^a$ on ${\cal M}$ and coordinates $x^i$ on ${\cal S}_0$, by
working in the coordinates $x^\mu$ adapted to the evolution. The
coordinate components of (\ref{eq:harm}) then take the form
\begin{equation}
  { \cal C}^\rho = 
  \frac{1}{2} g^{\mu\nu}g^{\rho\sigma}\left( \nablaz_{\mu} f_{\nu\sigma} 
     + \nablaz_{\mu} f_{\nu\sigma} - \nablaz_\sigma f_{\mu\nu} \right)
          =g^{\mu\nu}( \Gamma^\rho_{\mu\nu}- \Gammaz^\rho_{\mu\nu} ).
    \label{eq:charm}
\end{equation}
The requirement that $\Gamma^\rho:=g^{\mu\nu} \Gamma^\rho_{\mu\nu} =0$ is
the standard harmonic coordinate condition that $ \Box_g x^\mu  = 0$. In
the present case, setting ${\cal C}^\rho =0$ implies
$\Gamma^\rho=g^{\mu\nu} \Gammaz^\rho_{\mu\nu}$, which is an example of
harmonic coordinates with a forcing term, as discussed
in~\cite{Friedrich}.  (In numerical relativity, these have been called
generalized harmonic coordinates~\cite{pret}.) For more general forcing
terms the harmonic constraint takes the form
\begin{equation}
        \Gamma^\rho=H^\rho(x,g),
\label{eq:genharm}
\end{equation}
where restriction of the forcing term $H^\rho$ to depend only upon
$x^\mu$and  $g_{\mu\nu}$ ensures that the system remains well-posed.

In the adapted coordinates, the reduced Einstein equations
$E_{\mu\nu}=0$ form the desired quasilinear wave system for $f_{\mu\nu}$,
\begin{equation}
    g^{\rho\sigma}\nablaz_\rho\nablaz_\sigma f_{\mu\nu} 
 = 2 g_{\lambda\tau} g^{\rho\sigma} C^\lambda{}_{\mu \rho}C^\tau{}_{\nu\sigma}
   +4 C^\rho{}_{\sigma(\mu} g_{\nu)\lambda} 
    C^\lambda{}_{\rho \tau}g^{\sigma\tau}
    - 2 g^{\rho\sigma}\Rz^\lambda{}_{\rho\sigma(\mu} g_{\nu)\lambda} .
 \label{eq:beinst}
\end{equation}
With the hierarchy of Sommerfeld boundary conditions (\ref{eq:ak}) -
(\ref{eq:hl}), Theorem 1 of~\cite{isol} now applies and ensures that
(\ref{eq:beinst}) has a well-posed IBVP and, in particular, determines a
unique solution $f_{\mu\nu}$. In addition, the resulting metric
$g_{\mu\nu}$  must satisfy the harmonic constraints ${\cal C}^\rho=0$
because they are built into the initial data and boundary conditions.
(See Sec.~\ref{sec:applic} for details concerning constraint preservation.) 
Therefore $g_{\mu\nu}$ solves Einstein's equations.

The solution has been obtained in coordinates which are harmonic with
respect to the background $\gz_{\mu\nu}$, i.e. 
$g^{\mu\nu}(\Gamma^\rho_{\mu\nu}-\Gammaz^\rho_{\mu\nu})=0$. The resulting
spacetime metric $g_{ab}$ and background metric  $\gz_{ab}$ determined by
the data are in a gauge which depends upon the choice of evolution field
$t^a$. Under a diffeomorphism $\Psi$ of ${\cal M}$ which reduces to the
identity map on ${\cal S}_0$ and ${\cal T}$, we have
$(t^a, g_{ab},\gz_{ab}) \rightarrow (\Psi_* t^a, \Psi^* g_{ab}, \Psi^*
\gz_{ab} )$. Thus $\Psi^* g_{ab}$ is a diffeomorphic solution of
Einstein's equations which is harmonic with respect to the background
$\Psi^* \gz_{ab}$.

A more important question, which concerns the issue of {\it geometric
uniqueness} raised in~\cite{hjuerg}, is the behavior of the spacetime
under a diffeomorphism $\tilde \Psi$ of the disembodied boundary $\tilde
{\cal T}$. It is known for the Cauchy problem that data $(\tilde \Psi^*
\tilde h_{ab},\tilde \Psi^* \tilde k_{ab})$ on $\tilde {\cal S}_0$ lead
to a spacetime which is isometric to the spacetime with Cauchy data $(
\tilde h_{ab}, \tilde k_{ab})$. The analogous result holds for the
boundary data. Under a diffeomorphism $\tilde \Psi$ of $\tilde {\cal T}$
the boundary data maps according to
$(\tilde t, \tilde q , \tilde q^{ab},\tilde \sigma^{ab})  \rightarrow
(\tilde \Psi^* \tilde t,\tilde \Psi^* \tilde q ,  \tilde \Psi_* \tilde
q^{ab}, \tilde \Psi_* \tilde \sigma^{ab})$. After the embedding in ${\cal
M}$, this induces data 
$(\tilde \Psi^* t, \tilde \Psi^* \hat q ,  \tilde \Psi_* \hat q^{ab},
\tilde \Psi_* \hat \sigma^{ab})$ on ${\cal T}$. Now consider any smooth
extension of $\tilde \Psi$ to a diffeomorphism of ${\cal M}$. Suppose the
original data leads by the above construction to the spacetime metric
$g_{ab}$ with the evolution field $t^a$, background $\gz_{ab}$ and
relative geodesic curvature and relative shear $(q^a,\sigma^{ab})$ of the
outgoing null vector $K^a$ normal to the $t$-foliation of ${\cal T}$.
Then, by the same construction, the mapped data lead to the metric
$\tilde \Psi^* g_{ab}$ with the evolution field $\tilde \Psi_* t^a$,
background $\tilde \Psi^* \gz_{ab}$ and relative geodesic curvature and
relative shear $(\tilde \Psi_* q^a,\tilde \Psi_* \sigma^{ab})$ of the
outgoing null vector $\tilde \Psi_* K^a$ normal to the
$\tilde \Psi^* t$-foliation of ${\cal T}$. In this way, diffeomorphisms
of $\tilde {\cal S}$ and $\tilde {\cal T}$, which map their intersection
$\tilde {\cal B}_0$ into itself, generate an isometry class of vacuum
spacetimes.

We have thus shown that the resulting vacuum spacetimes are diffeomorphic
{\em if} the disembodied boundary data and initial Cauchy data are
diffeomorphic. This comprises a version of geometric uniqueness, which has
been pointed out as a missing ingredient in prior formulations of the
IBVP~\cite{hjuerg}. However, the result is not as strong as for  the pure
Cauchy problem for which the converse is also true: the resulting vacuum
spacetimes are diffeomorphic {\em only if} the initial data are
diffeomorphic. The converse is more complicated for the IBVP and does not
hold in the disembodied setting because the data $\tilde\sigma^{ab}$ are
superfluous. It is only the trace-free part of the embedded data
$\hat\sigma^{ab}$ which enters the shear $\sigma^{ab}$. But there is no way
to pick out the trace free part without knowledge of the metric, which is an
unknown at the stage of specifying data. (For linear perturbations, a
background metric could be used but this construction does not extend to
the nonlinear case.) This is perhaps an unavoidable feature of disembodied
Sommerfeld data for the IBVP.

As a result, disembodied boundary data which are not related by a diffeomorphism
can lead to isometric spacetimes. However, it follows by direct geometrical
construction that an isometry class of spacetimes does uniquely determine the
embedded boundary data $q^a$ and $\sigma^{ab}$ up to a diffeomorphism. It is
only upon the transition from $\sigma^{ab}$ to $\hat\sigma^{ab}$ that the
ambiguity of adding a trace enters. In this sense, a stronger version of
geometric uniqueness applies to the embedded data.

The Sommerfeld boundary conditions were based upon the background metric
obtained by the Lie transport (\ref{eq:gt0}) of the Cauchy data along the
streamlines of $t^a$. Modification of this transport law would lead to a
different background and the spacetime generated by the same disembodied data
would in general not be diffeomorphic. This is similar to the scalar wave
problem, for which the same boundary data would lead to different solutions if,
say, a Dirichlet boundary condition were used instead of a Sommerfeld condition.
In the present case, a new background metric implies a different form of the
Sommerfeld condition and, for fixed boundary data, the resulting spacetime will
(in general) not be isometric to the original spacetime. The effect of the
boundary data is changed by the difference between the two background
connections.

\section{Numerical application}
\label{sec:applic}

The motivation for this work has been the treatment of the outer boundary in
the numerical simulation of the inspiral and merger of binary black holes. The
boundary conditions (\ref{eq:ak}) - (\ref{eq:hl}), which lead to a strongly
well-posed IBVP, can be applied directly to any of the harmonic evolution
codes which have been used to simulate this binary
problem~\cite{fP05,fP06,lLmSlKrOoR06,oRlLmS07,bSdPlRjTjW07}. At present, none
of these harmonic codes incorporate boundary conditions that ensure strong
well-posedness. The closest example is the pseudo-spectral harmonic code
described in~\cite{lLmSlKrOoR06,oRlLmS07} which incorporates a second
differential order boundary condition which freezes the $\Psi_0$ Weyl
component and was shown to be well-posed in the generalized sense in the high
frequency limit~\cite{ruiz}.

An important attribute of strong well-posedness is the estimate of the
boundary values of the solution and its derivatives which are provided by an
energy conservation law obeyed by the principle part of the equations. This
boundary stability extends to the semi-discrete system of ordinary
differential equations in time which are obtained by replacing spatial
derivatives by finite differences obeying summation by parts (the discrete
counterpart of integration by parts), so that energy conservation caries over
to the semi-discrete problem. This stability then extends to the numerical
evolution algorithm obtained by applying an appropriate time integrator, such
as Runge-Kutta.

The advantage of boundary stability in numerical applications is that the
smoothness of the solution is unaffected by reflections off the boundary. This
avoids the long timescale instabilities which might otherwise arise from
multiple reflections. For the harmonic Einstein problem, each component of the
metric obeys a quasilinear wave equation so that it is straightforward to
develop a summation by parts algorithm based upon the standard energy
expression for a scalar wave in a curved spacetime~\cite{mBbSjW06}. A code
incorporating such an algorithm was applied, using a version of the well-posed
Sommerfeld boundary conditions presented here, to the test problem of a highly
nonlinear gauge wave propagating inside a cubic boundary~\cite{mBhKjW07}.
Although the proof of strong well-posedness given in~\cite{wpe} was based upon
a scalar wave energy differing by a small boost from the standard energy
expression, the successful results for this test problem confirm the
robustness of the underlying approach.

Strong well-posedness extends to more general quasi-harmonic formulations for
which the reduced Einstein equations have the form
\begin{equation}
     E^{\mu\nu}:= G^{\mu\nu} -\nabla^{(\mu}{\cal C}^{\nu)} 
       +\frac{1}{2}g^{\mu\nu}\nabla_\rho{\cal C}^\rho 
       +A^{\mu\nu}_\sigma {\cal C}^\sigma=0,
       \label{eq:creduced}
\end{equation}
where ${\cal C}^\rho= g^{\mu\nu}\Gamma^\rho_{\mu\nu} -H^\rho(x,g)$ are the 
generalized harmonic constraints (\ref{eq:genharm}) and the coefficients
$A^{\mu\nu}_\sigma$ have the dependence $A^{\mu\nu}_\sigma(x,g,\partial
g)$. (This includes constraint modified versions of the harmonic system.)
Constraint preservation follows from the Bianchi identity $\nabla_\mu
G^{\mu\nu} =0$ which implies a homogeneous wave equation for ${\cal
C}^\mu$,
\begin{equation}
    \nabla^\rho \nabla_\rho \, {\cal C}^\mu +R^\mu_\rho \,{\cal C}^\rho
  -2\nabla_\rho(A^{\mu\rho}_\sigma {\cal C}^\sigma) =0.
  \label{eq:bianchi}
\end{equation}
If the boundary conditions enforce ${\cal C}^\rho |_{\cal T}=0$ and the initial
data enforces  ${\cal C}^\rho |_{{\cal S}_0}=\partial_t {\cal C}^\rho |_{{\cal
S}_0}=0$ then the unique solution of (\ref{eq:bianchi}) is ${\cal C}^\rho=0$. As
a result, the Sommerfeld boundary conditions in the geometrical form
(\ref{eq:ak})  - (\ref{eq:sigqq}), along with (\ref{eq:hk}) - (\ref{eq:hl}) which
enforce ${\cal C}^\rho |_{\cal T}=0$,  lead to a well-posed harmonic IBVP in
which the harmonic constraints ${\cal C}^\rho=0$ are  satisfied everywhere. In
turn, (\ref{eq:creduced}) implies that the Hamiltonian and momentum constraints
$G^{\mu\nu}n_\nu =0$ are also satisfied.

Beyond the geometrical aspects of the harmonic IBVP, there are practical
concerns that arise in astrophysical applications. The linear time
dependence of the background metric (\ref{eq:g0}) could lead to deleterious
long time scale effects in numerical simulations. For that reason, it is
preferable to modify the prescription (\ref{eq:gt0}) for the background
metric so that (\ref{eq:g0}) changes to 
\begin{equation}
      {\gz}_{\mu\nu}=g_{\mu\nu}|_{t=0}
         +t e^{-\lambda t}(\partial_t g_{\mu\nu}|_{t=0})
     \label{eq:gm0}
\end{equation}
and the time dependence damps on a time scale determined by $\lambda$.
Alternatively, the purpose of tying the background metric to the initial Cauchy
data was to make clear that it did not affect the geometric uniqueness of the
solution. Otherwise, the Minkowski metric in the coordinates adapted to the
evolution could have been chosen as the background metric. This would leave
intact the well-posedness of the IBVP and might be the most expedient approach
in some numerical applications, although (\ref{eq:gm0}) has the advantage of
suppressing nonlinear effects in the early stage of an evolution.

Another practical concern is that the proper
boundary data $q^a$ and $\sigma$ are not normally known in an astrophysical
application. The practice in simulating an isolated system is to assume
homogeneous data on the artificial outer boundary, i.e. $q^a =\sigma=0$. 
Such homogeneous data in general leads to some spurious back reflection
from the boundary.  However,  as discussed in~\cite{isol} for the harmonic
IBVP, similar  boundary conditions on a round spherical outer boundary of
large surface area radius $R$  lead to reflected waves whose amplitude
falls off asymptotically as $1/R^3$ for both quadrupole gauge waves and
quadrupole gravitational waves. (The calculation assumes that the
linearized approximation is applicable. Modifications of the boundary
conditions by lower differential order terms involving factors of $R$ lead
to a faster $1/R^4$ falloff for the reflected quadrupole gravitational
waves.~\cite{isol})

The boundary conditions (\ref{eq:ak}) - (\ref{eq:hl}) are slightly
different than those considered in~\cite{isol} and, in particular, the
homogeneous boundary condition $\sigma=0$ leads to reflected waves with
only a $1/R^2$ falloff, as previously noted in~\cite{ruiz}. However,
improved performance can be obtained by taking advantage of the sensitivity
of the boundary conditions to the location of the indices. If
(\ref{eq:shdata}) and (\ref{eq:sigqq}) are replaced by 
\begin{equation}
       2\sigma= \frac{1}{2}Q^a Q^b(\nabla_a - \nablaz_a) K_b
      + \frac{1}{2}Q^a Q_b(\nabla_a - \nablaz_a) K^b
          = \frac {1}{2}(Q^b Q^c K^a
       - Q^b K^c Q^a) \nablaz_a f_{bc}   ,
        \label{eq:nshdata}
\end{equation}
then agreement with the corresponding boundary condition (94)
of~\cite{isol} is obtained. In that case, for the boundary conditions
(\ref{eq:ak}) - (\ref{eq:al}) and (\ref{eq:nshdata}), along with the
harmonic constraints (\ref{eq:hk}) - (\ref{eq:hl}), the amplitudes of the
reflected quadrupole gauge waves and quadrupole gravitational waves again
fall off asymptotically as $1/R^3$. Thus the application of these
Sommerfeld boundary conditions with homogeneous data results in
small back reflection from the boundary of an isolated system.

While the Sommerfeld boundary conditions considered here were developed for
the harmonic IBVP, their geometric nature allows them to be formally applied
to any metric version of the reduced Einstein equations, in particular the
alternative ``$3+1$'' formulations upon which much numerical work has been
based.  However, in that case, well-posedness and constraint preservation do
not necessarily follow. One approach to dealing with these issues would be to
re-express the $3+1$ formulations in terms of  the covariant 4-dimensional
$Z4$ theory~\cite{z4,z4sb}, in which  the reduced evolution equations take
the form  (\ref{eq:creduced}) but with the harmonic constraints ${\cal
C}^\mu$ replaced by a more general vector field $Z^\mu$, which can be used to
change the principle part of the resulting evolution system. The $Z4$
formalism has been shown to encompass the standard $3+1$ formulations,
including the Arnowitt-Deser-Misner (ADM)~\cite{adm}, the
Kidder-Scheel-Teukolsky~\cite{kst}, the Nagy-Ortiz-Reula~\cite{nor} and the
Baumgarte-Shapiro-Shibata-Nakamura (BSSN)~\cite{bssn1,bssn2} formulations. It
is possible that the close analogue between the $Z4$ and harmonic
formulations might be used to shed light on the analytic properties of $3+1$
systems. However, such an investigation of the well-posedenss of the $3+1$
IBVP is stymied by the fact that none of the systems used in numerical
applications have been shown to be symmetric hyperbolic. Some $3+1$ systems
have been show to be strongly hyperbolic, which ensures a well-posed Cauchy
problem, but symmetric hyperbolicity is required to apply the standard
theorems for a well-posed IBVP. For that reason, the  discussion here is
restricted to how the Sommerfeld boundary conditions developed here might be
applied in a manner consistent with what is presently known about $3+1$
formulations.

The boundary conditions (\ref{eq:ak}) - (\ref{eq:sigqq}) can be translated in a
straightforward way into a ``3+1'' decomposition $x^\mu =(t,x^i) =(t,x,y.z)$, in
which the metric takes the form  (\ref{eq:gdecom}) where, in terms of the lapse
and shift (\ref{eq:lapshif}),
\begin{equation}
    g_{tt} =-\alpha^2+ h_{ij}\beta^i \beta^j \, ,
      \quad  g_{ti} = h_{ij}\beta^j \, , \quad
          g_{ij} = h_{ij}.
\end{equation}
Although there is a different decomposition (\ref{eq:bdecom}) intrinsic to
the boundary, the outgoing null direction used in formulating a Sommerfeld
condition is intrinsic to both decompositions. Let $\hat N^\mu$ be the unit
normal to the boundary foliation ${\cal B}_t$, which lies in the $t=const$
Cauchy hypersurface, so that the metric (\ref{eq:gdecom}) has the further
decomposition
\begin{equation}
      g_{\mu\nu} = -n_\mu n_\nu +\hat N_\mu \hat N_\nu +Q_{\mu\nu},
      \label{eq:cdecom}
\end{equation}
where again  $Q_{\mu\nu}= Q_{(\mu}\bar Q _{\nu)}$ is the 2-metric intrinsic
to the foliation ${\cal B}_t$ of the boundary. Then the outgoing null
vector $K^\mu$ normal to the foliation has components $K^\mu =T^\mu +N^\mu
= e^{-\Theta} (n^\mu +\hat N^\mu)$  where $\Theta$ is the hyperbolic angle
arising from the velocity of the boundary relative to the Cauchy
hypersurfaces.  In terms of the lapse and normal component of the shift,
\begin{equation}
        \tanh\Theta = -\frac{1}{\alpha} \beta^i \hat N_i,
\end{equation}
where $\beta^i \hat N_i <0$ for a boundary which is moving inward
with respect to the Cauchy hypersurfaces. The sign of $\beta^i \hat N_i$
determines whether the advective derivative $n^\mu \partial_\mu$ is
outward ($\beta^i \hat N_i <0$) or inward ($\beta^i \hat N_i >0$)
at the boundary and, as a result, can affect the number
of required boundary conditions.

The Sommerfeld derivative takes the simple 3+1 form
\begin{equation}
   K^\mu\partial_\mu =e^{-\Theta}(n^\mu \partial_\mu
      +\hat N^\mu \partial_\mu), \quad 
         \quad  \hat N^\mu \partial_\mu = \hat N^i \partial_i,
\end{equation}
where $n^\mu \partial_\mu$ is the part containing the time derivative
$\partial_t$. Thus the Sommerfeld boundary conditions  (\ref{eq:ak}) -
(\ref{eq:sigqq}) can be used to supply boundary values for the time
derivatives of 6 metric components, or equivalently to supply boundary
values for 5 components of the extrinsic curvature $k_{\mu\nu}$ of the
Cauchy foliation and for the time derivative of one metric component. In
particular, (\ref{eq:ak}) - (\ref{eq:aq}) and (\ref{eq:sigqq}) supply
boundary values for $\hat N^\mu \hat N^\nu k_{\mu\nu}$,
$Q^\mu \hat N^\nu k_{\mu\nu}$ and $Q^\mu Q^\nu k_{\mu\nu}$ and
(\ref{eq:al}) supplies the boundary value for the time derivative of the
normal component of the shift $\hat N_i \partial_t \beta^i$, i.e. 
$(h^{zz})^{-1/2} \partial_t \beta^z$ for a boundary aligned with the
$z$-coordinate.

The remaining Sommerfeld boundary conditions (\ref{eq:hk}) - (\ref{eq:hl}),
which enforce the harmonic constraints, require modification depending upon the
particular $3+1$ formulation and gauge conditions. Only 6 components of
Einstein's equations are used in a $3+1$ evolution, with the gauge conditions
determining the evolution of the lapse and shift. As a first example, consider
the ADM formulation in which the 6 Einstein equations
\begin{equation}
       h_\mu^\rho h_\nu^\sigma R_{\rho\sigma}=0
      \label{eq:admev}
\end{equation}
are evolved. The evolution of the Hamiltonian and momentum
constraints $H:=G_{\mu\nu}n^\mu n^\nu$ and
$P^\mu:= h^{\mu\nu} n^\gamma G_{\nu\gamma}$ is governed by
the contracted Bianchi identity $\nabla_\nu G_\mu^\nu = 0$,
which gives rise to the symmetric hyperbolic constraint
propagation system
\begin{eqnarray}
     n^\gamma \partial_\gamma H - \partial_j P^j 
          &=& B^\gamma G_{\nu \gamma }n^\nu \\      
     n^\gamma \partial_\gamma P^i  - h^{ij} \partial_j H 
        &=& B^{i\gamma}  G_{\nu \gamma }n^\nu,
\end{eqnarray} 
where the coefficients $B^\gamma$ and $B^{\mu\gamma}$ arise from 
Christoffel symbols and do not enter the principle part.  An analysis of
this system shows that only one boundary condition is allowed provided
$\beta^i \hat N_i \le 0$, i.e provided the boundary is moving {\em inward}
relative to the Cauchy hypersurfaces. The theory of symmetric hyperbolic
systems then guarantees that all the constraints will be preserved if
\begin{equation}
     H+P^i N_i = G_{\mu\nu} n^\mu K^\nu =0
	   \label{eq:admcon} 
\end{equation}
is satisfied at the boundary.
(Additional boundary conditions are necessary for constraint preservation
if  $\beta^i \hat N_i > 0$.)
By combining the evolution system
(\ref{eq:admev}) with (\ref{eq:admcon}), this boundary condition
is equivalent to
\begin{equation}
           G_{\mu\nu} K^\mu K^\nu =0,
	   \label{eq:raych} 
\end{equation}
i.e. the Raychaudhuri equation~(cf. \cite{wald})
\begin{equation}
   K^\mu \partial _\mu \theta +\frac{1}{2}\theta^2+\sigma \bar\sigma =0,
\label{eq:drho}	
\end{equation}
where $\theta=Q^{\mu\nu}\nabla_\mu K_\nu$ is the expansion of the outgoing
null rays tangent to $K^\mu$. Thus, for the ADM system, constraint
preservation can be enforced by the Sommerfeld boundary condition
(\ref{eq:drho}) for $\theta$, which supplies the boundary values for the
remaining component $Q^{\mu\nu} k_{\mu\nu}$ of the extrinsic curvature.
Unfortunately, although the subsidiary constraint system is symmetric
hyperbolic, the ADM evolution system is only weakly hyperbolic and
consequently leads to unstable evolution.

In terms of astrophysical applications, the most important $3+1$
formulation is the BSSN system, which has been used by the majority of
groups~\cite{mCcLpMyZ06,jBjCdCmKjvM06b,pDfHdPeSeSrTjTjV06,jGetal07,bssnsom}
carrying out binary black hole and neutron star simulations. The
development of the BSSN formulation has proceeded through an
interplay between educated guesses and feedback from code performance. Only in
hindsight has its success spurred mathematical analysis, which has shown that
certain versions are strongly hyperbolic and thus have a well-posed Cauchy
problem~\cite{sartig,nor,gundl1}. Although significant progress has been made
in establishing some of the necessary conditions for well-posedness and
constraint preservation  of the IBVP~\cite{gundl2,horst,fritgom,nunsar}, 
there is still no satisfactory mathematical theory on which to base numerical
work. In current numerical practice, the boundary conditions for BSSN
evolution systems are applied in a naive, homogeneous Sommerfeld form to each
evolution variable (cf.~\cite{bssnsom}).

The geometric nature of the Sommerfeld boundary conditions (\ref{eq:ak}) -
(\ref{eq:sigqq}) and their role in a well-posed harmonic IBVP suggest that
they might lead to improved performance over the present boundary treatment
of the BSSN system. However, there are two complications. The first
involves the sign of the normal component of the shift $\beta^i \hat N_i $
at the boundary, which also entered the above discussion of the ADM
constraint system. A recent analysis~\cite{nunsar} of the BSSN evolution
system shows that the number of incoming fields at the boundary, and
therefore the number of required boundary conditions, depends upon the sign
of $ \beta^i \hat N_i $. A practical scheme for dealing with this would
require the Dirichlet boundary condition $ \beta^i \hat N_i =0$ (or some
similar Dirichlet condition to control the sign) rather than the Sommerfeld
condition (\ref{eq:al}) for the normal component of the shift.

The other complication for the BSSN system involves constraint
preservation. The BSSN evolution system enforces the 6 Einstein equations
\begin{equation}
      h_\mu^\rho h_\nu^\sigma R_{\rho\sigma}
      -\frac{2}{3}h_{\mu\nu} H=0,
      \label{eq:bssnmev}
\end{equation}
for which the constraint system implied by the Bianchi
identity takes the form
\begin{eqnarray}
     n^\gamma \partial_\gamma H 
     -\partial_j P^j  
          &=&  B^\gamma G_{\nu \gamma }n^\nu 
	 \nonumber \\      
     n^\gamma \partial_\gamma P^i 
      + \frac{1}{3} h^{ij} \partial_j H 
        &=&  B^{i\gamma}  G_{\nu \gamma }n^\nu.
	\label{eq:bssnmc}
\end{eqnarray}
This is no longer symmetric hyperbolic and would not lead to stable
constraint preservation even for the Cauchy problem. In order to remedy
this, the BSSN evolution system modifies (\ref{eq:bssnmev}) by mixing in a
set of auxiliary constraints, which combine with the constraint system
(\ref{eq:bssnmc}) to form a larger symmetric hyperbolic constraint system.
The freedom in the constraint-mixing parameters  and gauge conditions
complicates a general treatment. Here the discussion will be limited to a
particular choice~\cite{nunsar} for which the linearization off Minkowski
space leads to a symmetric hyperbolic system with a well-posed IBVP for the
case of a Dirichlet boundary condition $ \beta^i \hat N_i =0$ on the normal
component of the shift.  Although the nonlinear evolution system is no
longer symmetric hyperbolic, the boundary conditions for the linearized
theory  can be formally applied and lead to a symmetric hyperbolic
constraint system. Constraint preservation then follows for the parameter
range $(b_1 \le1,b_2 \le 1)$ in the boundary conditions given in equation
(97) of~\cite{nunsar}. The particular choice $b_1=0$, leads to the boundary
condition~\cite{pc}
\begin{equation}
         H-3  P^i N_i =G_{\mu\nu} n^\mu (n^\nu - 3N^\nu) = {\cal Z},
	   \label{eq:bssncon} 
\end{equation}
where ${\cal Z}$ represents contributions from the auxiliary constraints,
or, by using the evolution system (\ref{eq:bssnmev}),
\begin{equation}
           G_{\mu\nu} L^\mu L^\nu ={\cal Z}.
	   \label{eq:inraych} 
\end{equation}
It is a bizarre feature of the $3+1$ problem that the constraint preserving
boundary conditions switch from the outgoing Raychaudhuri form (\ref{eq:raych})
to the ingoing Raychaudhuri form (\ref{eq:inraych}) in going from the ADM to the
BSSN system. The Raychaudhuri equation for the outgoing null direction cannot be
imposed in the allowed range of $(b_1,b_2)$. Nevertheless, (\ref{eq:inraych})
can still be used to supply boundary values for the remaining
$Q^{\mu\nu} k_{\mu\nu}$ component of extrinsic curvature.

It appears from the above discussion that the formal application of the
Sommerfeld  boundary conditions to the BSSN system must be restricted to 
(\ref{eq:ak}), (\ref{eq:aq}) and (\ref{eq:sigqq}) which supply boundary
values for 5 components of the extrinsic curvature. Boundary values for the
remaining $Q^{\mu\nu} k_{\mu\nu}$ component must be obtained in accord with
constraint preservation, e.g. from (\ref{eq:inraych} ) or some variant
depending upon the particular formulation. The normal component of the
shift requires a Dirichlet boundary condition that determines its sign,
e.g.  $\beta^i N_i =0$. Appropriate boundary conditions for the lapse and
tangential components of the shift depend upon the specific gauge conditions
(see~\cite{nunsar} for an example). In the spirit of the BSSN formalism,
computational experiments would be necessary to determine whether (\ref{eq:ak}),
(\ref{eq:aq}) and (\ref{eq:sigqq}) lead to improved performance.

\section{Summary}

We have shown that a geometrically unique spacetime can be locally
constructed from Sommerfeld data determined by the fields
$(\tilde t,\tilde q,\tilde q^a,\tilde \sigma^{ab})$ on a disembodied
boundary  $\tilde {\cal T}$, along with the initial data prescribed on
${\tilde S}_0$. The boundary data specify a foliation $\tilde {\cal B}_t$
of $\tilde {\cal T}$ but involve no metric or other geometric properties
of the boundary. After the embedding of
$\tilde {\cal T}\cup {\tilde S}_0$ as the boundary
${\cal T}\cup {\cal S}_0$ of a 4-manifold ${\cal M}$, the induced fields
$( t,\hat q,\hat q^a,\hat \sigma^{ab})$ supply the necessary boundary
data for an isometry class of spacetime metrics which satisfy Einstein's
equations. Under diffeomorphisms of  the disembodied boundary
$\tilde {\cal T}$ and Cauchy hypersurface $\tilde {\cal S}_0$, the mapped
data determine diffeomorphic vacuum spacetimes.

The gauge in which a particular metric $g_{ab}$ is constructed via a
solution of the harmonic IBVP depends upon the choice of evolution field
$t^a$, which also determines an associated background metric $\gz_{ab}$
by Lie transport of the initial data.  Together $g_{ab}$ and $\gz_{ab}$ 
supply the geometric interpretation of the boundary data in terms of the
outgoing null vector $K^a$ normal to the foliation ${\cal B}_t$ of the
boundary. The field $q^a$, constructed from $\hat q$ and $\hat q^a$ via
(\ref{eq:qa}), is the geodesic curvature of $K^a$, relative to its
geodesic curvature computed with the background metric. The field
$\sigma^{ab}$ (or equivalently $\sigma$) computed from
$\hat \sigma^{ab}$ via (\ref{eq:shtendata}) (or (\ref{eq:shdata})) is the
shear of  $K^a$, relative to its background value. The resulting metric
is harmonic with respect to the background metric, according to
${\cal C}^\rho =0$ (see (\ref{eq:charm})). All possible choices of $t^a$
are related by diffeomorphism, so that all possible gauges are
included.

The Sommerfeld boundary conditions have direct application to harmonic
evolution codes used in simulating binary black holes, where they would
provide a numerical algorithm based upon a strongly well-posed IBVP.
Furthermore, for harmonic evolution, homogeneous Sommerfeld data gives
rise to asymptotically small back reflection of quadrupole waves from an
asymptotically large spherical outer boundary of an isolated system.

The geometric nature of the 6 boundary conditions (\ref{eq:ak}) -
(\ref{eq:sigqq}) suggests that they might also be applicable to codes
based upon a $3+1$ formulation, e.g. the BSSN system, as a way to reduce
spurious boundary effects generated by the naive Sommerfeld conditions
now in practice.  There are two caveats. First, strong well-posedness of
the IBVP does not directly apply to any present $3+1$ system. Second, the
additional Sommerfeld boundary conditions (\ref{eq:hk}) - (\ref{eq:hl})
only guarantee preservation of the Hamiltonian and momentum constraints
for harmonic  (or quasi-harmonic) formulations and would have to be
replaced in accord with constraint preservation. In the case of the particular
BSSN formulation shown in~\cite{nunsar} to posses a well-posed IBVP in the
linearized approximation, there are further complications. Instead of the
Sommerfeld condition (\ref{eq:al}), a Dirichlet boundary condition is required
to control the sign of the normal component of the shift. The 5 other Sommerfeld
conditions (\ref{eq:ak}), (\ref{eq:aq}) and (\ref{eq:sigqq}) can be used to
supply boundary values for 5 components of the extrinsic curvature. In addition,
the boundary values of the remaining component of extrinsic curvature can be
supplied by a constraint preserving boundary condition, but there is no apparent
way to do this in a Sommerfeld form. These complications would, at the least,
lead to more spurious reflection from the boundary than for a harmonic code. The
results of this paper can perhaps guide further experimentation and
investigation towards a modification of  the BSSN system that adds to its
successful features.

%%%%%%%%%%%%%%%%%%%%%%%%%%%%%%%%%%%%%%%%%%%%%%%%%%%%%%%%%%%%%%%%%%%%
\begin{acknowledgments}
%%%%%%%%%%%%%%%%%%%%%%%%%%%%%%%%%%%%%%%%%%%%%%%%%%%%%%%%%%%%%%%%%%%%

This research was supported by NSF grants PHY-0553597 and PHY-0854623 to
the University of Pittsburgh. Much of this work is based upon previous
collaborations with H-O. Kreiss, O. Reula and O. Sarbach and I have
benefited from their continued input. The motivation and guidance for the
new ideas developed here have come from discussions with H. Friedrich. I
am particular grateful to B. Schmidt for reading the manuscript and
suggesting improvements.

\end{acknowledgments}

\end{document}